\documentclass[aps,pra,twocolumn,showpacs,superscriptaddress]{revtex4-1}
\usepackage{amsmath,amssymb}
\usepackage{graphicx}
\usepackage[usenames]{color}
\newcommand{\op}[1]{\hat{#1}}
\newcommand{\dagop}[1]{\hat{#1}^{\dagger}}
\newcommand{\bo}[1]{{\mathbf{#1}}}

\newcommand{\wt}[1]{{\widetilde{#1}}}

\begin{document}

\setlength{\unitlength}{1cm}

\title{Tradeoffs for number-squeezing in collisions of Bose-Einstein condensates}

\author{P. Deuar}
\affiliation{Institute of Physics, Polish Academy of Sciences, Al. Lotnik\'ow
32/46, 02-668 Warsaw, Poland}

\author{T. Wasak}
\affiliation{Institute of Theoretical Physics, Physics Department, University
of Warsaw, Ho\.{z}a 69, PL-00-681 Warsaw, Poland}

\author{P. Zi\'n}
\affiliation{Narodowe Centrum Bada\'n J\c{a}drowych, ul. Ho\.za 69, 00-681 Warszawa, Poland}
\affiliation{Univ. Paris Sud, CNRS, LPTMS, UMR 8626, Orsay 91405 France}

\author{J. Chwede\'nczuk}
\affiliation{Institute of Theoretical Physics, Physics Department, University
of Warsaw, Ho\.{z}a 69, PL-00-681 Warsaw, Poland}

\author{M. Trippenbach}
\affiliation{Institute of Theoretical Physics, Physics Department, University
of Warsaw, Ho\.{z}a 69, PL-00-681 Warsaw, Poland}

\begin{abstract}
We investigate the factors that influence the usefulness of supersonic collisions of Bose-Einstein condensates as a potential source of entangled atomic pairs
by analyzing the reduction of the number difference fluctuations between regions of opposite momenta.
We show that non-monochromaticity of the mother clouds is typically the leading limitation on number squeezing, and that the squeezing becomes less robust to this effect as the density of pairs grows.
We develop a simple model that explains the relationship between density correlations and the number squeezing, allows one to estimate the squeezing from properties of the correlation peaks, and shows how the multi-mode nature of the scattering must be taken into account to understand the behavior of the pairing.
We analyze the impact of the Bose enhancement on the number squeezing, by introducing a simplified low-gain model. 
We conclude that as far as squeezing is concerned the preferable configuration occurs when atoms are scattered not uniformly but rather into two well separated regions.
\end{abstract}

\maketitle

\section{Introduction and Outline}
A supersonic collision of two Bose-Einstein condensates is a source of strongly correlated atomic pairs, which may be potentially used to
create spatially-separated entangled states of massive particles \cite{Jaskula10,Krachmalnicoff10,Vogels02,Perrin07,Deng99,Truscott,MetaStableReview,Kheruntsyan12}.
Such states could be used to extend the study of the Einstein-Podolsky-Rosen paradox \cite{horo,Reid09}, local realism \cite{kwiat} and Bell inequality tests \cite{Kofler12} 
into a regime where rest mass is non-negligible. In the context of quantum metrology, usefully entangled states allow one to surpass the Standard Quantum Limit
-- the maximum parameter estimation precision allowed by classical physics \cite{giovanetti,pezze}.

Scattering of atomic pairs can lead to reduced fluctuations of the relative population between two regions of opposite momenta. This effect, called the {\it number-squeezing}
and a closely related violation of the Cauchy-Schwartz inequality have been
recently observed in experiments \cite{Jaskula10,Bucker11,Kheruntsyan12,Bonneau12}.
Number squeezing, if accompanied by sufficiently high mutual coherence between the regions, is indicative of {\it spin squeezing} \cite{kita,wine}.
Spin squeezed states, which are known to be usefully entangled from a quantum metrology point of view \cite{pezze,ita1,wine}, have been recently engineered in a number of experiments
\cite{obert,app,gross,riedel,vule,chen}.
In addition, the perfectly number-squeezed ``twin-Fock'' state, which is not spin-squeezed but is nevertheless strongly entangled, has been recently observed by L\"ucke \emph{et al} \cite{smerzi}. 
Also pair-production setups were invented, where scattering is preferentially directed into only several spatial modes. These then contain more pairs per mode, and it is more convenient to bin them and possibly process further \cite{Kofler12}. In \cite{dall,pertot,Truscott} a four-wave-mixing type of process between two species of atoms was used to populate two counter-propagating clouds. In \cite{Bucker11} a BEC was transferred into the first excited state of a trapping potential and subsequent two-body collisions created a ``twin-beam'' system of correlated pairs. Finally, another approach used an optical lattice to allow correct phase-matching conditions into only a few selected modes\cite{Bonneau12}.

In some recent experiments, where atoms scatter into two well defined regions, a halo of overlapping spontaneously scattered modes 
and other imperfections such as spatial inhomogeneity of the mother clouds \cite{Ogren08,Ogren10} might limit the amount of number squeezing. 
An understanding of the main limitations and tradeoffs involved is important for future experiments.

We consider in detail the number difference squeezing between atoms with opposite momenta in a collision of two non-monochromatic BECs
\cite{MetaStableReview,Jaskula10}.
We simulate the collision using the positive-P method in the Bogoliubov approximation \cite{stab}.
We also introduce a simple yet intuitive model, which relates the number squeezing to the second order correlations between the scattered atoms and demonstrate
its validity in a wide range of parameters.
The main conclusion of this work is that the non-monochromatic nature of the mother clouds is the main limiting factor to strong number-squeezing in the scattering halo.
We argue that in the presence of many competing modes, smaller cloud densities are advantageous, because mode-mixing effects are destructive for the number-squeezing and they become more pronounced with higher cloud density.

The manuscript is organized as follows. In Section \ref{COLL} we introduce a model, which describes the creation of pairs in BECs collisions. We discuss the relevant physical parameters and develop an alternative low-gain model, useful for simulations when the bosonic amplification of pair scattering can be neglected.
In Section \ref{COH} we calculate the second order correlation function of the scattered atoms and explain how it consists of two parts -- co-linear and back-to-back momentum correlations, like in \cite{Molmer08}.
In Section \ref{ns-g2} we calculate the number-squeezing parameter and using a simple Gaussian model relate it to the second-order coherence of the system.
In Section \ref{NR} we present our numerical results
in the regime of both high- and low-density of the mother BECs and compare these results with the model. In the course of this analysis, the factors affecting the number squeezing become apparent.
We conclude in Section \ref{CONC}. Some technical details of the calculations and numerics presented in the main text are discussed in the Appendices.

\section{Theoretical model for BEC collision}
\label{COLL}

A single stationary BEC can be split into a superposition of two counter-propagating wave-packets by means of Bragg scattering \cite{Bragg}.
In a center-of-mass reference frame, the average velocity of each cloud is $\pm {\bf v}_{\rm rec}$ -- a recoil velocity due to an absorption and subsequent emission of a Bragg photon.
If the relative velocity $2v_{\rm rec}$ is approximately above the speed of sound $c=\sqrt{g\bar\rho/m}$, i.e. when the Mach number
$$
{\rm Ma}=2v_{\rm rec}/c>1,
$$
elastic collisions of particles from the two clouds lead to scattering of atomic pairs out of the BECs. Here $\bar\rho=N/(4/3\pi R_{\rm TF}^3)$ is the average density of an isotropic
condensate in the Thomas-Fermi approximation with radius $R_{\rm TF}$.

The dynamics of the scattering is governed by the energy- and momentum-conservation laws, $v_1^2+v_2^2\simeq2v_{\rm rec}^2$ and ${\bf v}_1+{\bf v}_2\simeq0$, where index $(1,2)$
labels the pair members. The equalities are only approximate, due to the momentum spread of the initial BEC and finite duration of the collision.
These conservation laws dictate that atoms are scattered onto a shell of radius $v_{\rm rec}$ centered around zero, called the scattering halo. This phenomenon was observed in many experiments
\cite{Jaskula10,Krachmalnicoff10,Vogels02,Perrin07,Deng99,Truscott,Kheruntsyan12}.
Atoms are usually registered after a long time of free expansion, when their positions approximately correspond to the momentum distribution just after the collision.

Since particles scatter in pairs, there is an expectation that the measured population difference between two opposite regions may fluctuate below the
shot-noise limit. In the idealized case of scattering from two plane waves, these fluctuations are suppressed down to zero, in analogy to the simplest model of
parametric down-conversion.
Our study takes on the task of generalizing this simple picture and calculating the number-squeezing in condensate collisions assuming a realistic shape and time evolution of the source, and including the time-dependent
interactions with the mean field after the scattering \cite{Deuar11epjd}.

\subsection{Collision parameters}

To focus on the essential features of the process, we consider the simplest case of the initial condensate prepared in the ground state of a spherical trap.
Depending on the non-linearity and the duration of the collision, the scattering of atoms can be either spontaneous or enter the Bose enhanced regime. Therefore,
we introduce a dimensionless parameter
$$
\gamma=t_{\rm col}/t_{\rm nl}
$$
 to distinguish between these two possible scenarios. Here, $t_{\rm col}=R_{\rm TF}/v_{\rm rec}$
is the duration of the scattering process, while $t_{\rm nl}=\hbar/(\frac 13g\bar\rho)$ is the rate of the two-body collisions.
The one-third in the denominator approximately accounts for the fact that collisions between the atoms take place mostly in the center of the trap, where the density is high.
It has been demonstrated \cite{Zin05,Zin06,Chwedenczuk06,Chwedenczuk08,Deuar11epjd} that when $\gamma\gtrsim1$, the system enters the stimulated regime.

\begin{figure}[htb]
  \hspace*{-0.45cm}\includegraphics[clip]{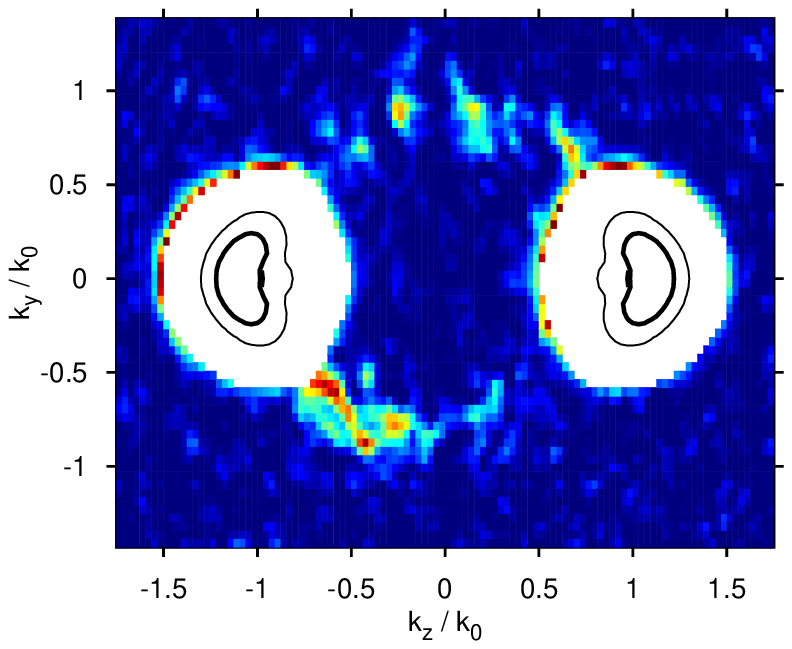}\\
  \includegraphics[clip]{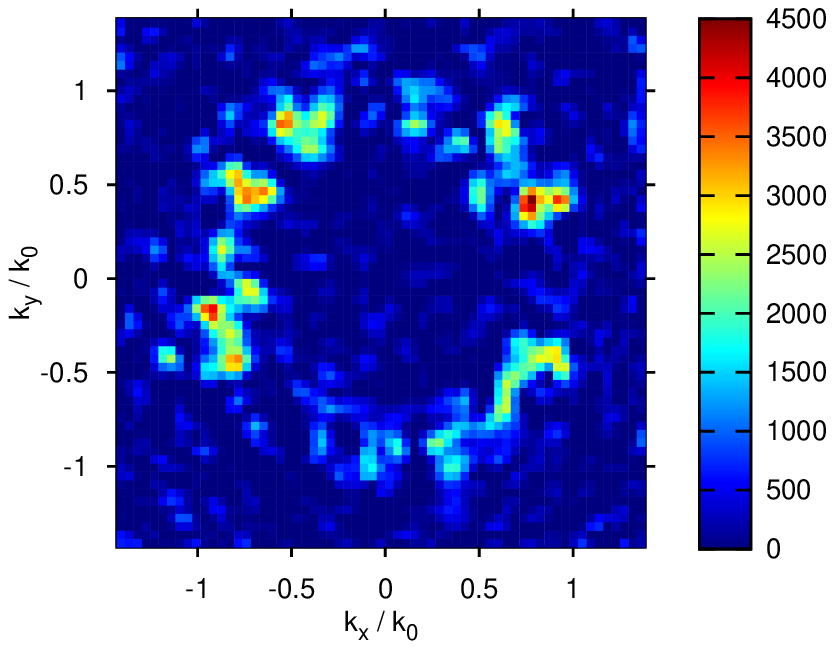}
  \caption{(Color online) Outcome of a single realization of a collision of two BECs in the dense $\gamma=1.02$ case (In a simulation of (\ref{STABeq},\ref{GPeq})\,). Shown are cross-sections through the density of atoms after the end of the collision ($t=1.6t_{\rm coll}$),
in the $k_z-k_y$ momentum plane (top), and $k_x-k_y$ plane (bottom).
    Clearly, a halo of scattered atoms is forming, nevertheless the atoms at the opposite regions of the halo are poorly
    aligned, which implies the absence of number squeezing in the system.
Data scaled to units of $[k_0]{}^{-3}$.
The white areas are saturated at this colour scale, the bold (thin) black contours are at $0.1$ ($0.01$) of the peak condensate density.
  \label{f1}}
\end{figure}

We perform numerical simulations using  parameters of metastable $^4$He atoms, i.e. $m=6.65\times10^{-27}$~kg, $a_s=7.5\times10^{-9}$~m and take
$v_{\rm rec}=10$~cm/s. To address both spontaneous and stimulated regimes we consider the following cases.
In the first, which we call the {\it dilute gas case}, the sample consists of $N=17\,540$ atoms in the initial BEC,
and $\omega=2\pi\times928~{\rm s}^{-1}$. Here ${\rm Ma}=13.12$ and $\gamma=0.24$, hence the scattering is spontaneous all along.
In the opposite {\it dense gas case}, we take $N=74\,360$ and $\omega=2\pi\times1911~{\rm s}^{-1}$, which gives ${\rm Ma}=6.56$
and $\gamma=1.02$, so the Bosonic enhancement becomes significant.
In both cases, we simulate the scattering process until $t=1.7t_{\rm coll}$ -- a time at which the collision is completed.
In Fig.~\ref{f1}, cuts through the halo density are shown for the dense case at the end of the collision. The result was obtained using the positive-P numerical method, which is discussed
in detail below.
Note that although the \textit{ensemble average} of the momentum distribution is symmetric around zero,
isolated density grains are present in a single realization, and they show an evident and massive lack of this symmetry.
This is related to \textit{increased} fluctuations of the population imbalance between regions with opposite momentum, and will be investigated further below.

\subsection{BEC wave-function}

We model the initial trapped BEC by a c-number wave-function $\phi_0(\bo{x})$  which is a solution of the stationary Gross-Pitaevskii equation,
\begin{equation}\label{GPeqSt}
  \mu\,\phi_0(\bo{x}) = \left[-\frac{\hbar^2}{2m}\nabla^2 + V_{\rm trap}(\bo{x})+ g|\phi_0(\bo{x})|^2\right]\phi_0(\bo{x}).
\end{equation}
Here $m$ is the atomic mass, $\mu$ is the chemical potential, $g=4\pi\hbar^2a_s/m$ the interaction strength related to s-wave scattering length $a_s$,
and $V_{\rm trap}(\bo{x})=\frac12 m\omega^2\bo{x}^2$ is the harmonic trapping potential with a frequency $\omega$.
The collision is triggered by a pair of brief Bragg pulses shined onto the condensate \cite{Bragg}.
Consequently, a superposition of two counter-propagating, mutually coherent atomic clouds is created
\begin{equation}\label{GPic}
  \phi(\bo{x},t=0) = \phi_0(\bo{x})\frac{e^{ik_0 z}+e^{-ik_0 z}}{\sqrt 2}\equiv\phi_{+}({\bf x})+\phi_{-}({\bf x}),
\end{equation}
where $k_0=mv_{\rm rec}/\hbar$ is the wave-vector associated with the recoil velocity.
After the pulses are applied, the trap is switched off, and the two condensates begin to move apart, activating the collision process.

\subsection{Positive-P method}

To describe the scattered atoms we introduce a bosonic operator $\hat\delta(\bo{x,t})$.
In the spirit of the Bogoliubov approximation, we use the linearized equations of motion for the field $\hat\delta$,
assuming that the number of scattered atoms is a small fraction of the condensate population and the self-interaction of $\op{\delta}$ can be neglected,
\begin{eqnarray} \label{Bog}
  i \hbar\,\partial_t\op{\delta}(\bo{x},t)&=&\left[ -\frac{\hbar^2}{2m}\nabla^2+ 2g|\phi(\bo{x},t)|^2\right] \op{\delta}(\bo{x},t)\nonumber\\
  &+&g\phi^2(\bo{x},t) \op{\delta}^\dagger(\bo{x},t),
\end{eqnarray}
The coherent mean field part $\phi(\bo{x},t)$ during the collision evolves according to the time-dependent Gross-Pitaevskii equation,
\begin{equation}\label{GPeq}
  i\hbar\,\partial_t\phi(\bo{x},t)= \left[-\frac{\hbar^2}{2m}\nabla^2 + g|\phi(\bo{x},t)|^2\right]\phi(\bo{x},t).
\end{equation}

To solve the coupled equations (\ref{Bog}) and (\ref{GPeq}), we use the stochastic positive-P method \cite{stab},
where instead of directly solving Eq.~(\ref{Bog}) for $\op{\delta}$ we sample the distribution of two complex fields $\psi(\bo{x},t)$ and $\wt{\psi}(\bo{x},t)$.
The Bogoliubov equation (\ref{Bog}) corresponds to a pair of stochastic It\={o} equations,
\begin{subequations}\label{STABeq}
  \begin{eqnarray}
    i\hbar\,\partial_t\psi({\bo{x}},t)&=&\left(-\frac{\hbar^2}{2m} \nabla^2 + 2g|\phi({\bo{x}},t)|^2\right)\psi({\bo{x}},t)\label{STABpsia}\\
    &&\hspace*{-0.9cm}+g\phi^2({\bo{x}},t)\wt{\psi}({\bo{x}},t)^*+\sqrt{i\hbar g}\,\phi({\bo{x}},t) \xi({\bo{x}},t), \label{STABpsib}\\
    i\hbar\,\partial_t\wt{\psi}({\bo{x}},t)&=&\left(-\frac{\hbar^2}{2m} \nabla^2 + 2g|\phi({\bo{x}},t)|^2\right)\wt{\psi}({\bo{x}},t)\label{STABpsita}\\
    &&\hspace*{-0.9cm}+g\phi^2({\bo{x}},t)\psi({\bo{x}},t)^*+\sqrt{i\hbar g}\,\phi({\bo{x}},t) \wt{\xi}({\bo{x}},t).\label{STABpsitb}
  \end{eqnarray}
\end{subequations}

Here $\xi({\bo{x}},t)$ and $\wt{\xi}({\bo{x}},t)$ are independent, real stochastic Gaussian noise fields with zero mean and second moments given by
$\langle \xi({\bo{x}},t)\wt{\xi}({\bo{x}}',t') \rangle =0$ and

\begin{equation}
  \langle \xi({\bo{x}},t)\xi({\bo{x}}',t') \rangle
  = \langle \wt{\xi}({\bo{x}},t)\wt{\xi}({\bo{x}}',t') \rangle
  =\delta({\bo{x}}-{\bo{x}}')\delta(t-t').
\end{equation}
Within the stochastic positive-P method, any physical quantity can be obtained with the mapping  $\op{\delta}\to\psi$ and $\dagop{\delta}\to\wt{\psi}^*$, and replacing quantum averages
of normally-ordered operators by stochastic averages \cite{Deuar02}.
Equations (\ref{STABeq}) recover the exact quantum dynamics of Eq.~(\ref{Bog}) in the limit of an infinite number of samples.

\subsection{Scattering in the absence of bosonic enhancement}
\label{NOBE-TH}

The main goal of this study is to investigate how the number squeezing between two regions of the halo is affected by various phenomena which occur during the collision.
Various phenomena influence the dynamics in a complicated way and it seems very advantageous if we could, at least theoretically,``turn on/off'' some of them to isolate their effects.
For instance, the impact of the BEC mean-field on the scattered particles can be easily ``controlled'' by including or excluding the second term of the
right-hand-sides in lines (\ref{STABpsia}) and (\ref{STABpsita}). Similarly, the mean-field repulsion in the evolution of the source BEC can be controlled by setting $g=0$ by hand in
Eq.~(\ref{GPeq}). One can further simplify the dynamics by modeling the two counter-propagating condensates with non-expanding Gaussians, substituted for $\phi({\bf x},t)$ into Eqs
(\ref{STABeq}).

In this subsection we propose a simple perturbative model, which describes the condensate collision in the absence of bosonic stimulation.
We start by introducing a hierarchy of fields
\begin{equation}\label{exp}
  \op{\delta}(\mathbf{x},t) =\sum_{j=0}^\infty \op{\delta}^{(j)}(\mathbf{x},t).
\end{equation}
The lowest order term $\op{\delta}^{(0)}(\mathbf{x},t)$ is the solution of the free equation (i.e. without the additional particle creation term in the Bogoliubov field)
\begin{equation}\label{loword}
  i\hbar\,\partial_t\op{\delta}^{(0)}(\mathbf{x},t)=H_0(\mathbf{x},t)\op{\delta}^{(0)}(\mathbf{x},t),
\end{equation}
where $H_0(\mathbf{x},t)=-\frac{\hbar^2}{2m}\nabla^2+ 2g|\phi(\bo{x},t)|^2$. The higher terms of expansion (\ref{exp}) evolve according to
\begin{equation}\label{nobe}
  i\hbar\,\partial_t\op{\delta}^{(j)}(\mathbf{x},t)=H_0(\mathbf{x},t)\op{\delta}^{(j)}(\mathbf{x},t) + g\phi^2(\bo{x},t)\op{\delta}^{(j-1)\dagger}(\mathbf{r},t).
\end{equation}
In this approach, the bosonic enhancement appears only in the higher order fields  and can be excluded by restricting the dynamics only to the two lowest ones, namely $j=0$ and $j=1$.

As we argue in \cite{pert_Wasak}, the set of two coupled equations for $j=0$ and $j=1$ can be formally solved by replacing the operators with complex stochastic fields, i.e.
$\op{\delta}^{(j)}(\mathbf{x},t) \to \delta^{(j)}(\mathbf{x},t)$ and $\op{\delta}^{(j)\dagger}(\mathbf{x},t) \to \delta^{(j)}(\mathbf{x},t)^*$.
Then, the c-number equivalent of (\ref{loword}) and (\ref{nobe}) is solved numerically from the initial conditions consisting of $\delta^{(1)}(\mathbf{x},0)=0$ and setting $\delta^{(0)}(\mathbf{x},0)$
as a random complex Gaussian field with zero mean and the variances
$\langle \delta^{(0)}(\mathbf{x},0)^* \delta^{(0)}(\mathbf{x'},0)\rangle = \delta(\mathbf{x}-\mathbf{x}')$ and $\langle \delta^{(0)}(\mathbf{x},0) \delta^{(0)}(\mathbf{x'},0)\rangle = 0$.
It is important to note that -- contrary to the positive-P equations (\ref{STABeq}) -- the stochasticity is introduced only through the initial conditions, very much like in the truncated Wigner method, but with twice the variance.

With the solutions $\delta^{(0)}(\mathbf{x},t)$ and $\delta^{(1)}(\mathbf{x},t)$ at hand, one can reproduce the observables. For instance the lowest order correlation function
is given by
\begin{equation}
\langle \op{\delta}^{\dagger}(\mathbf{x},t)\op{\delta}(\mathbf{x}',t) \rangle = \overline{\delta^{(1)}(\mathbf{x},t)^*\delta^{(1)}(\mathbf{x}',t)},
\end{equation}
where the over-bar denotes averaging over the ensemble of initial conditions.

In the following Section, we analyze some formal properties of the Bogoliubov equation (\ref{Bog}) and introduce the second order correlation function of scattered atoms,
which, as will be argued in Section \ref{ns-g2}, determines the amount of number-squeezing in the system.

\section{Two-body correlations of scattered atoms}
\label{COH}

The normalized second order correlation function of scattered atoms in momentum space is defined as
\begin{equation}\label{g2_gen}
  g^{(2)}({\bf k},{\bf k}')=\frac{\langle\dagop{\delta}(\bo{k}) \dagop{\delta}(\bo{k}') \op{\delta}(\bo{k}') \op{\delta}(\bo{k})\rangle}
  {\langle\dagop{\delta}(\bo{k})\op{\delta}(\bo{k})\rangle\langle\dagop{\delta}(\bo{k}') \op{\delta}(\bo{k}')\rangle}.
\end{equation}
Henceforth, we omit the explicit time dependence from the operator $\op{\delta}(\bo{k},t)$.
Note that since the Bogoliubov equation of motion~(\ref{Bog}) is linear and the initial state of scattered atoms is a vacuum, with the help of Wick's theorem we can write
\begin{equation}\label{g2}
  g^{(2)}({\bf k},{\bf k}')=1+\frac{|G^{(1)}({\bf k},{\bf k}')|^2+|M({\bf k},{\bf k}')|^2}{G^{(1)}({\bf k},{\bf k})G^{(1)}({\bf k}',{\bf k}')}.
\end{equation}
Here, $G^{(1)}$ is the one-body density matrix of the scattered atoms defined as
\begin{equation}
  G^{(1)}({\bf k},{\bf k}')=\langle\dagop{\delta}(\bo{k})\op{\delta}(\bo{k}')\rangle,
\end{equation}
and $M$,  the anomalous density \cite{Zin06,Chwedenczuk08}, is
\begin{equation}
  M({\bf k},{\bf k}')=\langle\op{\delta}(\bo{k})\op{\delta}(\bo{k}')\rangle.
\end{equation}

In order to find a natural interpretation of the components of (\ref{g2}) we make some further simplifications in our model, that are used only in this Section.
First, consider the following simplified model \cite{Zin05,Zin06,Chwedenczuk08} of the collision dynamics, described by the equation
\begin{equation}\label{bog_sim}
  i\hbar\,\partial_t\op{\delta}(\bo{x},t)=-\frac{\hbar^2\nabla^2 }{2m}\op{\delta}(\bo{x},t)+2g\phi_+(\bo{x},t)\phi_-(\bo{x},t)\op{\delta}^\dagger(\bo{x},t),
\end{equation}
Here, a pair of atoms is taken from counter-propagating condensates $\phi_{\pm}$ -- defined in Eq.~(\ref{GPic}) --
and placed in the field of scattered atoms. As compared with Eq.~(\ref{Bog}),
this model neglects the impact of the mean field of the colliding BECs on the scattering process, as well as terms proportional to $\phi_{\pm}^2$ that are strongly non energy-conserving in the halo.

Now let us make a second simplification regarding the internal dynamics of the two colliding wave-packets. In general the two functions $\phi_{\pm}({\bf x},t)$ evolve according to the Eq.~(\ref{GPeq}), but for the sake of the present considerations we neglect the non-linear term and use the equations of motion of free expansion.
In this case, $\phi_{\pm}({\bf k},t)=\phi_{\pm}({\bf k})e^{-i\frac{\hbar k^2}{2m}t}$.
Such a ``reduced Bogoliubov'' model with these two approximations has been widely used previously to investigate the dynamics of the collision\cite{Bach02,Yurovsky02,Zin05,Chwedenczuk06,Zin06,Chwedenczuk08,Trippenbach00,Band00,Band01,Chwedenczuk04,Ogren09}, and was investigated in some detail in \cite{Deuar11epjd}.

Equation (\ref{bog_sim}) can be solved up to the first order in the perturbative regime, to obtain at times long after the collision
\begin{eqnarray}
  \label{ma}
  M({\bf k},{\bf k}')&=&\frac{m g}{i \pi^2 \hbar^2}\int\!\int\!\!d^3{\bf k}_1 d^3{\bf k}_2\,\phi_{-}({\bf k}_1)\phi_{+}({\bf k}_2) \times\label{em}\\
  & & \times\delta(k_1^2+k_2^2-k^2-k'^2)\delta^{(3)}({\bf k}_1+{\bf k}_2-{\bf k}-{\bf k}').\nonumber
\end{eqnarray}
The anomalous density $M({\bf k},{\bf k}')$ can be interpreted as the probability amplitude for having one atom with momentum ${\bf k}$ and the second with ${\bf k}'$.
These two momenta come from a coherent superposition of probability amplitudes describing elementary collision events (energy and momentum conservation laws are satisfied) between atoms from the BECs that come with probability amplitudes $\phi_\pm$.
Since the condensate functions $\phi_{\pm}$ in~(\ref{em}) are localized around $\pm k_0$, the anomalous density $M({\bf k},{\bf k}')$ is non-vanishing only when
${\bf k}\simeq-{\bf k}'$ and $|{\bf k}|\simeq k_0$. In this sense, $M$ describes scattered atomic pairs, correlated for opposite momenta. Related arguments have been presented in \cite{Molmer08}.

Using similar arguments, we obtain a useful relation that is valid within Bogoliubov theory in the perturbative regime
\begin{equation}\label{rho}
  G^{(1)}({\bf k},{\bf k}')=\int\! d^3{\bf k}''M^*({\bf k},{\bf k}'')M({\bf k}'',{\bf k}').
\end{equation}
As we argued above, the first anomalous density gives the contribution to the integral when ${\bf k}\simeq-{\bf k}''$ and the second when ${\bf k}''\simeq-{\bf k}'$,
hence the one-body density matrix is non-zero only if ${\bf k}\simeq{\bf k}'$ and $|{\bf k}|\simeq k_0$.
In conclusion, the scattered atoms are described either by the {\it co-linear} part for ${\bf k}\simeq{\bf k}'$,
\begin{equation}\label{cl}
  g^{(2)}_{\rm cl}({\bf k},{\bf k}')=g^{(2)}({\bf k},{\bf k}'\approx{\bf k}) \approx
  1+\frac{|G^{(1)}({\bf k},{\bf k}')|^2}{G^{(1)}({\bf k},{\bf k})G^{(1)}({\bf k}',{\bf k}')}
\end{equation}
or the {\it back-to-back} part when their wave-vectors are almost opposite ${\bf k}\simeq-{\bf k}'$,
\begin{equation}\label{bb}
  g^{(2)}_{\rm bb}({\bf k},{\bf k}')=g^{(2)}({\bf k},{\bf k}'\approx-{\bf k}) \approx
  1+\frac{|M({\bf k},{\bf k}')|^2}{G^{(1)}({\bf k},{\bf k})G^{(1)}({\bf k}',{\bf k}')}.
\end{equation}

In our numerical simulations we have seen that the above
interpretation of $M$ and $G^{(1)}$ is valid to a high degree of accuracy for a wide range of parameters, even when the assumptions introduced above are not fully valid.
Therefore, throughout the rest this work, we use the division of the second order correlation function into
the co-linear (\ref{cl}) and back-to-back part (\ref{bb}).

At this stage we are ready to introduce the number-squeezing parameter and show how it relates to the $g^{(2)}$ correlation function in various relevant regimes.

\section{Number squeezing in a multi-mode system}
\label{ns-g2}

In this section we define the number-squeezing parameter and show how it is related to the second-order coherence of the system.

Atoms are registered (and counted) in two \emph{bins}, say $a$ and $b$, encompassing volumes $V_{a/b}$ in momentum space. The corresponding atom-number operators read
\begin{equation}\label{num_many}
  \hat n_{a/b}=\int\limits_{V_{a/b}}\!\!\!\! d^3\bo{k}\ \dagop{\delta}(\bo{k}) \op{\delta}(\bo{k}).
\end{equation}
We introduce the number-difference operator $\hat n=\hat n_a-\hat n_b$ and define the number-squeezing parameter as follows
\begin{equation}\label{etaG}
  \eta^2=\frac{\Delta^2\hat n}{\bar n},
\end{equation}
where $\bar n=\langle\hat n_a\rangle+\langle\hat n_b\rangle$ is the total number of particles in both bins.
States that have sub-Poissonian population imbalance fluctuations $\eta^2<1$ are called number squeezed.
In the symmetric case $\langle\hat n_a\rangle=\langle\hat n_b\rangle$, number squeezing
is equivalent to violation of the Cauchy-Schwartz inequality \cite{Kheruntsyan12}, which implies the presence of non-classical correlations in the system.

Using Eq.~(\ref{num_many}) and the definition of $g^{(2)}$ from Eq.~(\ref{g2_gen}), we obtain that
\begin{equation}\label{eta2avg}
  \eta^2 = 1+\frac{G_{aa}+G_{bb}-2G_{ab}}{\bar n},
\end{equation}
where the $G_{ij}$ stands for a two-fold integral
\begin{equation}
  G_{ij}=
  \int\limits_{{\cal V}_i}\!\! d^3\bo{k}\!\int\limits_{{\cal V}_j}\!\! d^3\bo{k}'\,g^{(2)}(\mathbf{k},\mathbf{k}')\,n(\mathbf{k})\,n(\mathbf{k}').\label{int_G}
\end{equation}
Here $n(\mathbf{k})=\langle \dagop{\delta}(\bo{k}) \op{\delta}(\bo{k})\rangle$.
Note that if the two regions $a$ and $b$ are located on the opposite sides of the halo, $G_{aa}$ and $G_{bb}$ depend on the co-linear correlation function $g^{(2)}_{\rm cl}$,
while $G_{ab}$ is a functional of $g^{(2)}_{\rm bb}$.

To handle the integrals (\ref{int_G}) and evaluate the number-squeezing parameter $\eta^2$ for typical situations, let us model the normalized co-linear and back-to-back correlation functions by Gaussians
\begin{equation}
  g_{\rm cl}^{(2)}(\mathbf{k},\mathbf{k}')=1+h_{\rm cl}\!\!\!\prod_{i=z,t,r}\!\!\!e^{-\frac{(k_i-k_i')^2}{2(\sigma^{\rm cl}_i)^2}},
\label{g2clgaus}\end{equation}
for $\mathbf{k}\simeq\mathbf{k}'$ and
\begin{equation}
  g_{\rm bb}^{(2)}(\mathbf{k},\mathbf{k}')=1+h_{\rm bb}\!\!\!\prod_{i=z,t,r}\!\!\!e^{-\frac{(k_i+k_i')^2}{2(\sigma^{\rm bb}_i)^2}},\label{g2bbgaus}
\end{equation}
for $\mathbf{k}\simeq-\mathbf{k}'$.
This way we only need to extract the amplitudes $h_{\rm cl/bb}$ and the widths $\sigma^{\rm cl/bb}_i$ from the results of the numerical simulations, similarly
to the analysis of the Cauchy-Schwartz violation in\cite{Kheruntsyan12}. The Gaussian form is a very good match to the calculated and also the experimental correlation shapes.

\begin{figure}[hbt]
  \includegraphics[clip,scale=0.33]{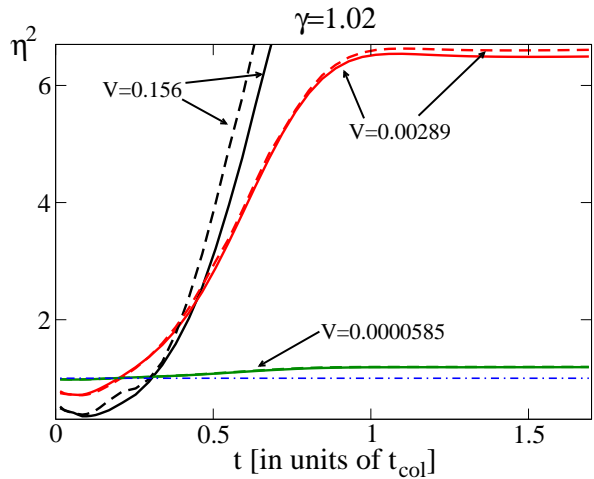}
  \caption{(Color online) The number squeezing parameter $\eta^2$ in the dense $\gamma=1.02$ case as a function of time for three different bin volumes given in $k_0^3$ units. The solid lines are from a numerical solution
    of the full model (\ref{STABeq},\ref{GPeq}). The dashed lines are predictions of the Gaussian correlation model (\ref{eta2best}). The dash-dotted horizontal line denotes the shot-noise level $\eta^2=1$.
  }\label{lN_t_vs_mod}
\end{figure}

The product runs over three orthogonal directions where the axis $z$ is along the collision direction,
while $r$ and $t$ are orthogonal to each other and lie in the $x-y$ plane, corresponding to radial ($r$) and tangential ($t$) directions. Additionally, we model the density $n({\bo k})$ in (\ref{int_G}) in the following way. We assume that the bin widths $L_z$ and $L_t$ (in the $z$ and $t$ directions respectively) are small enough for the density to be practically constant. This assumption is in our case well satisfied. On the other hand, the density quickly decays in the
 $r$ direction. To account for this drop relatively simply, we model the density in this direction with a step function, centered around the peak of the halo. The width $w_r$ is deduced from a Gaussian fit to the radial profile of the halo density obtained numerically.

As shown in the Appendix \ref{APP-eta}, the assumptions introduced above lead to the approximate expressions
\begin{equation}\label{eta2best}
  \eta^2 = 1 + \frac{\bar{n}}2\Big(h_{\rm cl} f_z^{\rm cl}f_t^{\rm cl}f_r^{\rm cl} - h_{\rm bb}f_z^{\rm bb}f_t^{\rm bb}f_r^{\rm bb}\Big).
\end{equation}
Here we have introduced the function
\begin{equation}\label{Gexpr3}
  f^{\rm cl/bb}_i= \frac{1}{u_i^2}\left[u_i\sqrt{\frac{\pi}{2}}\ {\rm erf}(u_i\sqrt{2}) - \frac{1}{2}\left(1-e^{-2u_i^2}\right) \right]
\end{equation}
and the normalized bin widths are $u^{\rm cl/bb}_{z,t} = L_{z,t}\Big/2\sigma^{\rm cl/bb}_{z,t}$ in the $z,t$ directions, while in the radial direction the limited density manifests itself by
$u^{\rm cl/bb}_r ={\rm min}(L_r,w_r)\Big/2\sigma^{\rm cl/bb}_{r}$. The limiting behavior of $f$ is $1-u^2/3$ for small bins $u$ much narrower than the correlations, and $(1/u)\sqrt{\pi/2}$ for large bins $u\to\infty$.
The above parametrization is convenient, though elaborate. However, when bin widths tend to infinity, we obtain a particularly useful expression
\begin{equation}\label{Vestsig}
  \eta^2=1+(2\pi)^{\frac32}\frac{\bar n}{2V}\left[h_{\rm cl}\sigma^{\rm cl}_z\sigma^{\rm cl}_t\sigma^{\rm cl}_r-h_{\rm bb}\sigma^{\rm bb}_z\sigma^{\rm bb}_t\sigma^{\rm bb}_r\right].
\end{equation}
Hence, in general, in the multi-mode system, number squeezing depends on the number of particles in the bins $\bar n$, but also depends on the correlation amplitudes $h_i$ and the widths of the correlation functions $\sigma_i^{\rm cl/bb}$. The situation simplifies dramatically when the bins are very small. In this case we get
\begin{equation}\label{gen_eta}
  \eta^2 = 1 +\frac{\bar n}2\left(h_{\rm cl} - h_{\rm bb}\right),
\end{equation}
and the quantum state is number-squeezed as long as $h_{\rm cl}<h_{\rm bb}$.

To place this in context, an additional important remark is in order. If we consider a pure two-mode pair production model governed by
the Hamiltonian $\op{H}= \op{a}\op{b} + \dagop{a}\dagop{b}$ we obtain simply $\eta^2\equiv0$. Therefore we see
that in a multi-mode system one cannot use the intuitions from the two-mode model to predict such quantities as the number-squeezing parameter,
even when the bins being considered are very small.

\begin{figure}[hbt]
  \includegraphics[clip,scale=0.33]{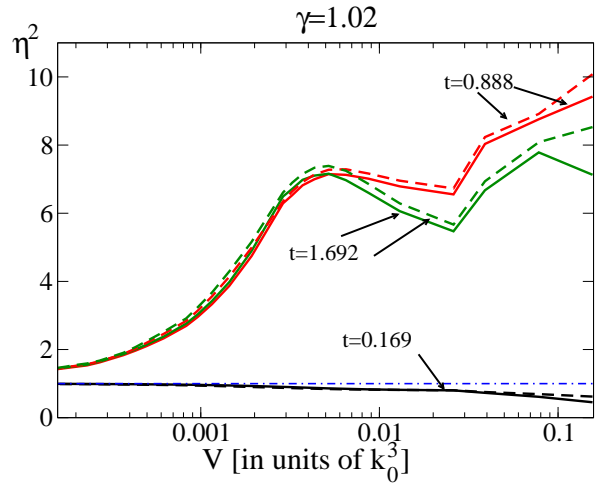}
  \caption{(Color online) The number squeezing parameter $\eta^2$ in the dense $\gamma=1.02$ case as a function of bin volume (log scale) calculated at
    three different scattering times (in units of $t_\mathrm{col}$).
    The solid lines are from a numerical solution of the full model (\ref{STABeq},\ref{GPeq}). The dashed lines are predictions of the Gaussian correlation model (\ref{eta2best}). The dash-dotted horizontal line denotes the shot-noise level $\eta^2=1$.
  }\label{lN_k_vs_mod}
\end{figure}

\section{Simulation results and analysis}
\label{NR}

In this Section we perform a systematic study of the number squeezing parameter in condensate collisions and identify the physical phenomena, which have the largest impact on $\eta^2$.

\subsection{$\gamma=1.02$ case}

We begin with the case of a dense mother cloud, with $\gamma=1.02$, so that the bosonic enhancement comes into play at some stage during the collision. In Fig.~\ref{lN_t_vs_mod}
we plot $\eta^2$ as a function of time for three different bin volumes $V$ (for details how the bins are chosen, refer to Appendix \ref{APP-BIN}). The solid curves, which are a
result of the simulation of the full positive-P Equations (\ref{STABeq}) are compared with the analytical prediction (\ref{eta2best}) based on correlation properties. The latter requires both the height and the widths
of the second order correlation function as input, and these are extracted from the simulation. The Figure shows very good agreement between that model and a ``direct'' evaluation of the number squeezing
by counting the number of particles in the two bins. Note that apart from early times, $\eta^2>1$ so opposite bins do not reveal number squeezing.

To check to what extent the number squeezing is a result of a particular choice of bins, in Fig.~\ref{lN_k_vs_mod} we plot $\eta^2$ as a function of the bin volume $V$ at three times. Again, the agreement between the model (\ref{eta2best}) and the simulation is very good. This Figure confirms that the absence of number squeezing in the scattering halo is a general feature regardless of binning details.

\begin{figure}[hbt]
  \includegraphics[clip,scale=0.33]{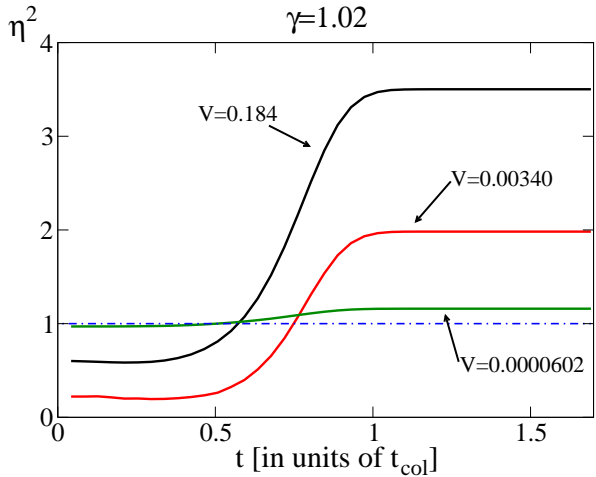}
  \caption{(Color online) The number squeezing parameter $\eta^2$ in the dense $\gamma=1.02$ case as a function of time for three different bin volumes (in $k_0^3$ units), as predicted by the reduced Bogoliubov model (RBM) (\ref{bog_sim},\ref{rbmGP}).
    The dash-dotted horizontal line denotes the shot-noise level $\eta^2=1$.
  }\label{lN_t_vs_rbm}
\end{figure}

In the past, it has been conjectured that the quantum correlations in the halo in such experiments are degraded by interactions with the mean field, or due to the time-variation fo the source cloud.
To identify which process is responsible for such dramatic loss of number squeezing at later times, we first compare the results of the full positive-P method with a maximally reduced Bogoliubov Method (RBM).
The evolution of the colliding condensates is simplified to a counter-propagation of the two initial clouds with fixed shape, and additionally, in equations (\ref{STABeq}) we include only the pair production term, so it simplifies to (\ref{bog_sim}). Both the mean field self-interaction of the BECs and its impact on the scattered particles are neglected. Free kinetic dispersion is also suppressed, so that the equations of motion for the condensate field are
\begin{equation}\label{rbmGP}
i\hbar\partial_t\phi_{\pm}(\mathbf{x},t) = \frac{\hbar^2k_0}{2m}(\mp2i\partial_x-k_0)\phi_{\pm}(\bo{x},t).
\end{equation}
 The condensates do not spread and the scattering process is maximally simplified. Figure \ref{lN_t_vs_rbm} shows the number squeezing parameter as a function of time as predicted by the RBM. Although $\eta^2$ does not grow as strongly as in Fig.~\ref{lN_t_vs_mod}, still the atom-difference fluctuations surpass the shot-noise limit.
Therefore, it is \emph{neither} the mean-field interaction nor the spreading of the BECs that have the major impact on the number squeezing parameter.

\begin{figure}[hbt]
  \includegraphics[clip,scale=0.33]{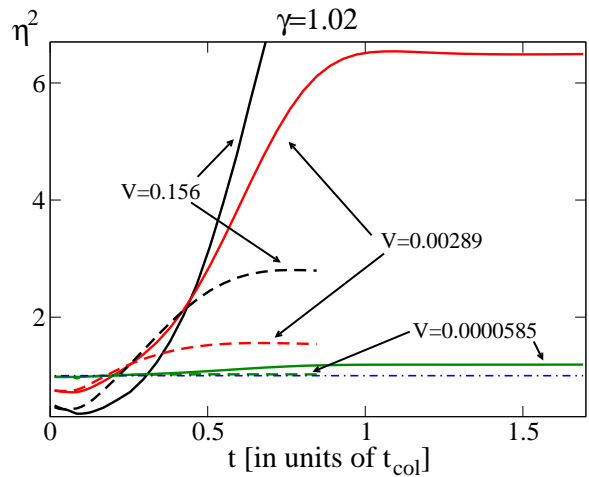}
  \caption{(Color online) The number squeezing parameter $\eta^2$ in the dense $\gamma=1.02$ case as a function of time for three different bin volumes given in $k_0^3$ units. The solid lines are from a numerical solution
    of the full model (\ref{STABeq},\ref{GPeq}). The dashed lines are predictions of the model without Bose enhancement (\ref{loword},\ref{nobe}). The dash-dotted horizontal line denotes the shot-noise level $\eta^2=1$.}
  \label{lN_t_vs_nobe}
\end{figure}

Next, we simulate the condensate collision using the numerical method which \textit{a priori} excludes the bosonic enhancement, introduced in Sec.~(\ref{NOBE-TH}), see Eq.~(\ref{nobe}). In Fig.~\ref{lN_t_vs_nobe} we compare the results obtained in the full positive-P simulation and the non-enhanced method. We plot $\eta^2$ as a function of time for three different bin volumes. Although the growth of the number squeezing parameter
is less violent in the absence of bosonic enhancement, still $\eta^2$ is above the shot-noise limit. As expected, for short times $t\lesssim0.1$ the outcomes of the two methods agree very well -- the system is still in the spontaneous regime.

\begin{figure}[hbt]
  \includegraphics[clip,scale=0.33]{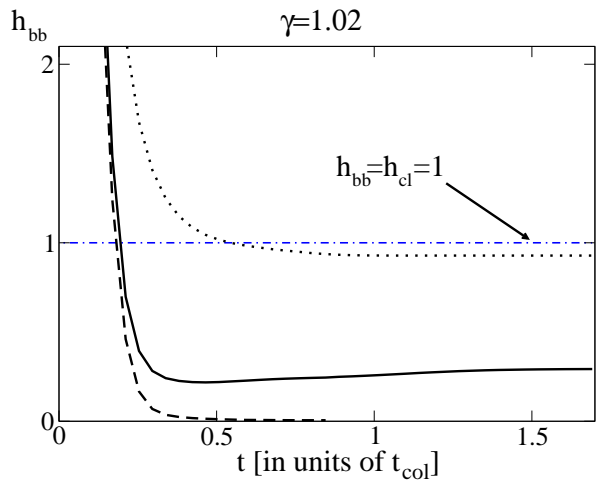}
  \caption{(Color online) The peak height of the back-to-back correlation function $h_{bb}$ as a function of time for the dense case $\gamma=1.02$. The solid line is from a numerical solution of the full model
 (\ref{STABeq},\ref{GPeq}), the dashed line from the model without Bose enhancement (\ref{loword},\ref{nobe}), while the dotted line comes from the reduced Bogoliubov model (RBM) (\ref{bog_sim},\ref{rbmGP}). The dot-dashed line shows the border value of $h_{\rm bb}=1$ when the back-to-back and collinear peaks are equal, and small-bin number-squeezing disappears.}\label{PHight}
\end{figure}

Finally, in Fig.~\ref{PHight} we draw the peak value of the back-to-back correlation function, namely the $h_{bb}$ defined in Eq.~(\ref{g2bbgaus}). As indicated by equations (\ref{eta2best}) and (\ref{Vestsig}),
the number squeezing is more likely to occur for high $h_{bb}$, the widths of the correlation functions being the other factor. We see that in all three methods (full positive-P, RBM, and non-enhanced) give $h_{bb}<h_{cl}\equiv1$ at long times, a limit below which  (\ref{gen_eta}) indicates that small bins can never be number-squeezed.

\subsection{$\gamma=0.24$ case}

\begin{figure}[hbt]
  \includegraphics[clip,scale=0.33]{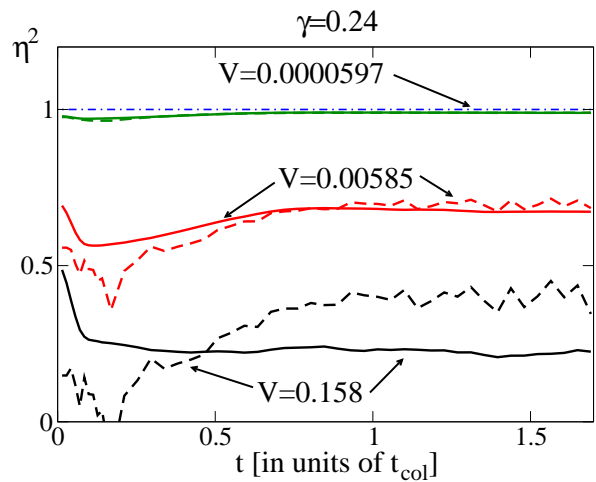}
  \caption{(Color online) The number squeezing parameter $\eta^2$ in the dilute $\gamma=0.24$ case as a function of time for three different bin volumes given in $k_0^3$ units. The solid lines come from a numerical solution
    of the full model (\ref{STABeq},\ref{GPeq}). The dashed lines are predictions of the Gaussian correlation model (\ref{eta2best}). The dash-dotted horizontal line denotes the shot-noise level $\eta^2=1$.
  }\label{sN_t_vs_mod}
\end{figure}

Here we investigate the number squeezing parameter in the dilute case, when $\gamma=0.24$ and the bosonic enhancement is absent. First, in Fig.~\ref{sN_t_vs_mod} we plot
$\eta^2$ as a function of time resulting from the positive-P method (\ref{STABeq}) and from the Gaussian correlation model (\ref{eta2best}). The agreement is satisfactory, although the predictions of the model are very noisy.
This is a result of the small number of scattered atoms. When the signal is low, it is difficult to extract the widths and peak values of the correlation functions that enter into the model. Nevertheless, we observe a major difference between the dense and the dilute case. In the latter, the bins are number-squeezed, irrespectively of their volume and the time.

\begin{figure}[hbt]
  \includegraphics[clip,scale=0.33]{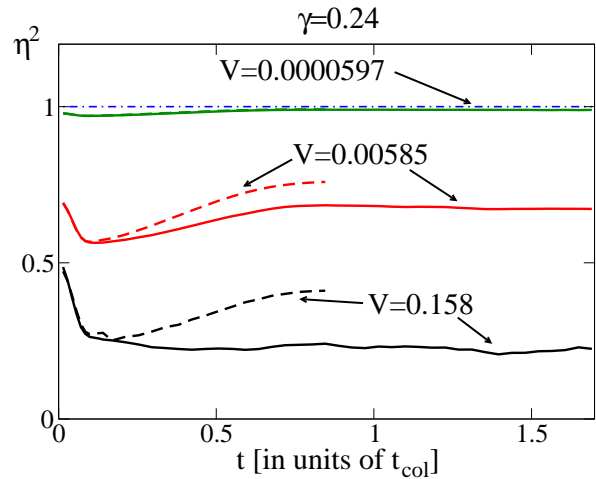}
  \caption{(Color online) The number squeezing parameter $\eta^2$ in the dilute $\gamma=0.24$ case as a function of time for three different bin volumes given in $k_0^3$ units. The solid lines result from a numerical solution
    of the full model (\ref{STABeq},\ref{GPeq}). The dashed lines are predictions of the model without Bose enhancement (\ref{loword},\ref{nobe}). The dash-dotted horizontal line denotes the shot-noise level $\eta^2=1$.
  }\label{sN_t_vs_nobe}
\end{figure}

We confirm that the system is indeed in the spontaneous regime, by comparing in Fig.~\ref{sN_t_vs_nobe} the number squeezing parameter as predicted by equations (\ref{STABeq})
and the non-enhanced method (\ref{nobe}). We do not observe any major discrepancy between these two results, and conclude that the system is indeed in the low-gain regime.

\begin{figure}[hbt]
  \includegraphics[clip,scale=0.33]{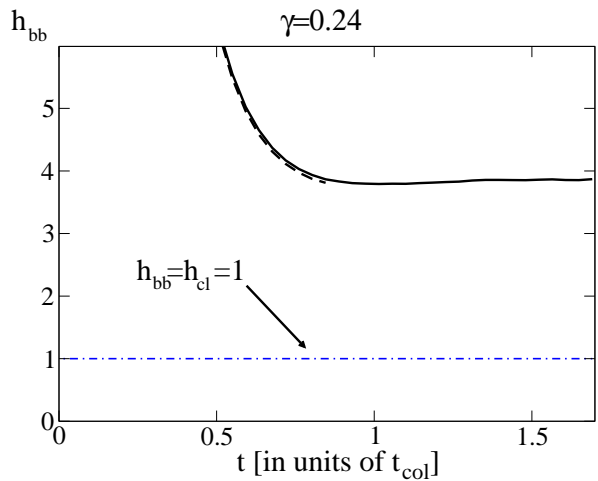}
  \caption{(Color online) The peak height of the back-to-back correlation function $h_{bb}$ as a function of time for the dilute case $\gamma=0.24$. The solid line is from a numerical solution of the full model
 (\ref{STABeq},\ref{GPeq}), the dashed line from the model without Bose enhancement (\ref{loword},\ref{nobe}). The dot-dashed line denotes the border value of  $h_{\rm bb}=1$
    when the back-to-back and collinear peaks are equal.
  }\label{sN_t_hbb}
\end{figure}

\begin{figure}[hbt]
  \includegraphics[clip,scale=0.33]{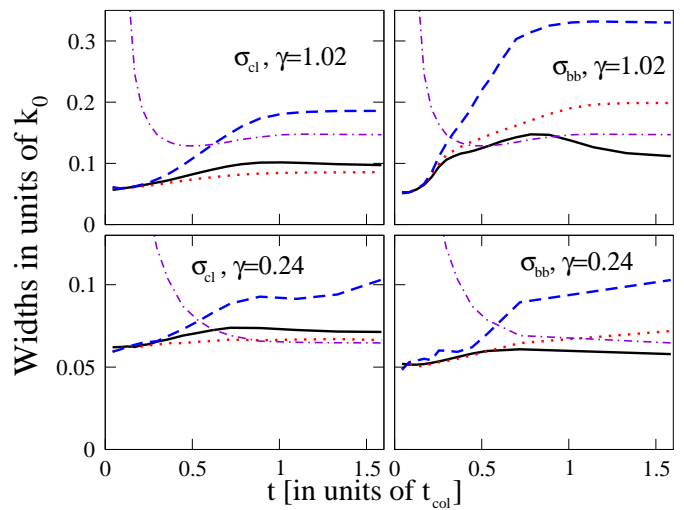}
  \caption{(Color online) The widths of the co-linear (left column) and back-to-back (right column) correlation functions for the dense (top row) and dilute case (bottom row) as a function of time. Fitted to the numerical solution
    of the full model (\ref{STABeq},\ref{GPeq}).
    The solid black line is the width in the axial direction $\sigma_z$,
    the dotted red line is in the tangential direction to the halo center $\sigma_t$, while the dashed blue line is in the radial direction $\sigma_r$.
    The dot-dashed violet line is the Gaussian fitted half-width of the halo density $w_r$, which narrows in time due to energy-time uncertainty principle. All widths are given in $k_0$ units.
  }\label{sigma}
\end{figure}

To understand why the number squeezing is present in the $\gamma=0.24$ case, in Fig.~\ref{sN_t_hbb}
we plot the peak of the back-to-back correlation function as a function of time. We see, that it is
far from reaching the border value of $h_{bb}=h_{cl}\equiv1$. Also, we compare this value with the one predicted by the non-enhanced method (\ref{nobe}) and find excellent agreement.
The number squeezing parameter depends not only on the peak values of the correlation functions, but also on their widths. In Fig.~\ref{sigma} we compare the widths of the
correlation function for $\gamma=1.02$ and $\gamma=0.24$ as a function of time. We notice that in both situations the back-to-back and co-linear widths are comparable to each other.
In the dense case, the back-to-back widths are slightly larger than the co-linear, which should favor $\eta^2<1$ for large bins, as indicated by Eq.~(\ref{Vestsig}).

The lack of number squeezing for any bin size for long times of the $\gamma=1.02$ case, must be therefore attributed to the drop of the peak height $h_{bb}$.
This drop is due to the non-monochromatic nature of the parent BECs. Their momentum spread leads to a non-zero width of the back-to-back correlation function,
which in turn results in scattering into not exactly opposite momentum modes. The pair of atoms can end up in non-opposite bins.
Nevertheless, when the number of scattered atoms is low -- as in the dilute gas case or at early times in the dense gas case -- there is a high probability of finding
a single or a few {\it correlated} pairs in the opposite regions, which is related to a high value of $h_{bb}$. Crucially, the probability of having some {\it uncorrelated} pairs in the opposite bin is low, because
the neighboring bins are mostly empty.
However, when the number of scattered particles grows, the chance that neighboring bins are empty becomes small, and uncorrelated atoms spill over into the opposite bin.
In this way, the $\eta^2$ fluctuations grow, since there is a significant amount of {\it uncorrelated} pairs in the opposite bins.
Figure \ref{f1}, which is an outcome of a single collision in the dense gas case, is an illustration of this scenario. There are some clearly visible regions, where the atoms form a single
large speckle, while on the other side of the halo two distinct speckles are present.

\subsection{Collision of two plane-waves}

\begin{figure}[hbt]
  \includegraphics[clip,scale=0.33]{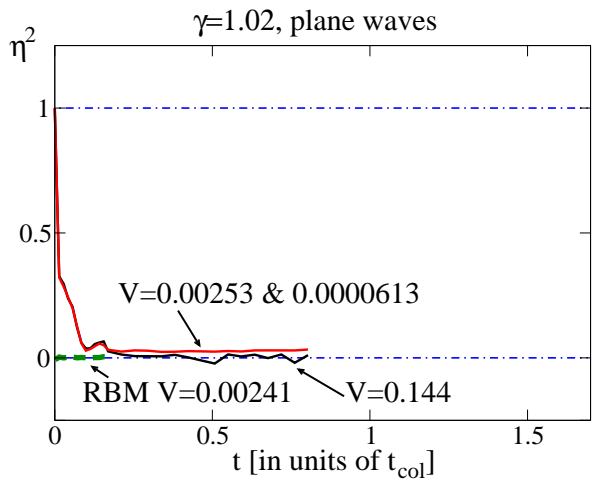}
  \caption{(Color online) The number squeezing parameter $\eta^2$ in the dense $\gamma=1.02$ case as a function of time for several different bin volumes given in $k_0^3$ units. Here,
    the  BECs are replaced with the plane-waves.
    The solid lines come from a numerical solution
    of the full model (\ref{STABeq},\ref{GPeq}), while the dashed green line from a reduced Bogoliubov model (RBM) simulation (\ref{bog_sim},\ref{rbmGP}).
    The dash-dotted horizontal lines denote the shot-noise and zero levels.
  }\label{plane_eta}
\end{figure}

To confirm the conjecture formulated in the previous paragraph,
we simulate a collision of two monochromatic plane waves. We use the same parameters as in the $\gamma=1.02$ dense case, but replace the initial BECs with
monochromatic peaks in momentum space at $\mathbf{k}=(0,0,\pm k_0)$ (and replace $|\phi|^2$ in the evolution equation (\ref{GPeq}) with the mean density). This mean density is chosen equal to the peak density in the usual $\gamma=1.02$ case, i.e. $(10/7)\bar\rho$.
In Fig.~\ref{plane_eta} we plot $\eta^2$ as a function of time for three different bin volumes.
In all cases, the number squeezing is near perfect ($\eta^2\approx0.03$), despite a huge total number of scattered atoms  $\sim10^7$.
The residual slightly nonzero value of $\eta^2$ is induced by the presence of a sea of short-lived particles from virtual scattering events. The RBM, which does not include this
non-resonant effect, gives $\eta^2=0$ within statistical accuracy.

\begin{figure}[hbt]
  \includegraphics[clip,scale=0.33]{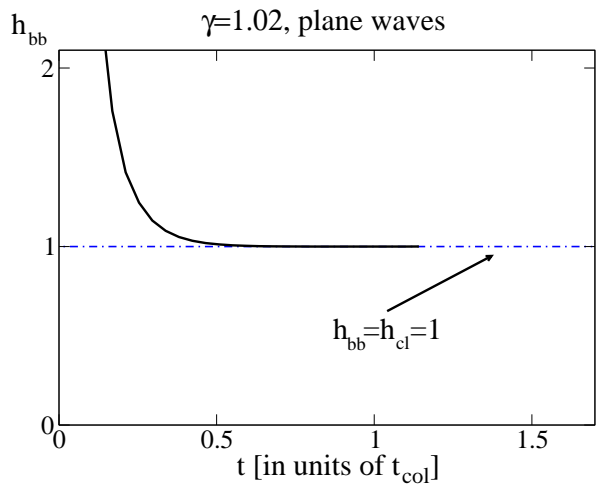}
  \caption{(Color online) The peak height of the back-to-back correlation function $h_{bb}$ as a function of time for the dense case $\gamma=1.02$ and two colliding plane-waves. The solid line is from a numerical solution of the full model
 (\ref{STABeq},\ref{GPeq}), while dot-dashed line denotes the border value of  $h_{\rm bb}=1$ when the back-to-back and collinear peaks are equal.
  }\label{plane_hbb}
\end{figure}

We also plot the peak height of the back-to-back function as a function of time, see Fig.~\ref{plane_hbb}.
At long times it becomes indistinguishable from the border value of unity, but at this stage the number of atoms in the halo and the bin occupation $\bar{n}$ is very large. However, from the model (\ref{gen_eta}) one sees that the minimal, fully squeezed value of $\eta^2=0$ corresponds to $h_{\rm bb} = 1+2/\bar{n}$, which is extremely close to unity, so that this remains consistent with the observed strong number squeezing in Fig.~\ref{plane_eta}.

\section{Conclusions}
\label{CONC}

We have performed a systematic study of the number squeezing parameter between two regions of opposite momenta in the scattering halo formed during collisions of two 
BECs. We have shown that the number squeezing depends strongly on the bin size, density of the mother clouds, mean-field interactions, and above all on the spectral purity of the mother clouds. In the dilute case, the number squeezing is evident, since the number of scattered pairs is low. Therefore, once a single atom is detected at momentum ${\bf k}$, there is a high probability of finding one (and just one) at ${\bf k}'\approx-{\bf k}$. Such a setup can be useful for investigation of the foundations of quantum mechanics, and indeed most Bell inequality tests in quantum optics have been carried out with weak sources where the granularity of the boson field becomes visible. 

On the other hand,  having only a single -- or a few -- correlated pairs is unattractive from a quantum interferometry point of view. The dense, or ``squeezing'' \cite{Duan00} regime is preferable. However, we find here that
 when the amount of scattered atoms is high, $\eta^2$ grows rapidly. By performing additional simulations with plane-waves instead of Gaussian condensates, we have related this behavior to the non-monochromaticity of the colliding BECs.
As the sources are not point-like in momentum space and the back-to-back correlation function has a non-zero range, the scattered pairs are not perfectly aligned. This in turn
results in imperfect correlation between opposite bins, which is additionally amplified by the increased number of stray atoms that enter them from neighboring regions. The non-monochromatic nature of the source clouds appears then to be the effect that underlies most of the degradation of pair correlations and number squeezing in the halo.

Our results have potential application for the setup and analysis of existing and future experiments for the production of correlated atomic pairs from ultracold atom sources, 
as these are almost always multi-mode, i.e. non-monochromatic to an appreciable degree. 
Quasicondensate and 1D phase-fluctuation aspects can be tamed by an appropriate choice of the counting bins \cite{Bucker11,Bonneau12,pertot,Dennis10,Haine11}. 
However, additional \emph{broad} spontaneous halos are observed 
in some mode-selective experiments, including directed-beam \cite{Bonneau12},  dressed state \cite{Williams12}, and four-wave mixing setups \cite{Vogels02,dall}.

The effect of the width of the mother cloud on short-time behavior was analyzed by Ogren and Kheruntsyan for BEC collisions\cite{Ogren09} 
and molecular dissociation\cite{Ogren08,Ogren10}, with improvement of squeezing as the cloud broadens. 
An important point we demonstrate here is that while the short-time behavior can be squeezed both at low and high densities, any squeezing is lost if the number of particles grows too far.
Thus, as clouds become denser, the onus on achieving or maintaining monochromaticity grows. Conversely, if squeezing is lost for a given geometry, 
it should be recoverable if the density is reduced sufficiently so that a significant proportion of measurements are free of stray unpaired atoms.

Preferable conditions for number squeezing are satisfied when atoms scatter into well separated regions, because the pool of atoms that go into a broad halo is strongly reduced, 
and the likelihood of unpaired stray atoms in the measuring bins is lessened. 
Such conditions have been demonstrated recently in twin beam experiments \cite{Bucker11}, four-wave mixing experiments with two-component matter waves \cite{dall,Truscott,pertot}, 
or with an optical lattice that selects only a pair of phase-matched modes \cite{Bonneau12} by modifying the dispersion relation.

\begin{acknowledgments}
  P.D. acknowledges the support of the Polish Government grant 1697/7PRUE/2010/7 and of the EU grant PERG06-GA-2009-256291.
  T.W. acknowledges the Foundation for Polish Science International Ph.D. Projects Programme co-financed by the EU European Regional Development Fund.
  J.Ch. acknowledges the Foundation for Polish Science International TEAM Programme co-financed by the EU European Regional Development Fund.
  T.W. and J.Ch. were supported by the National Science Center grant no. DEC-2011/03/D/ST2/00200.
  M.T. was supported by the National Science Center grant.
\end{acknowledgments}

\appendix

\section{Division of the halo into bins}
\label{APP-BIN}

To calculate the number-squeezing parameter we follow a similar procedure to that used in recent experiments \cite{Jaskula10,Kheruntsyan12}.
We take an annular ``washer-shaped'' region matched to the position and width of the halo, that excludes regions near the condensates, and divide it into zones. The matched dimensions of the washer-shaped region in the various
calculations are shown in Table~\ref{TableDim}.

\begin{table}
\begin{tabular}{|c|c|c|c|c|}
\hline
Figures													& $\gamma$	& small 	& large	& maximum	\\
														& 		& radius	& radius	& $|k_z|$	\\
\hline\hline
\ref{lN_t_vs_mod},\ref{lN_k_vs_mod},\ref{lN_t_vs_nobe},\ref{PHight},\ref{f7} 		& 1.02	& 0.671	& 1.029	& 0.333	\\
\ref{lN_t_vs_rbm},\ref{PHight} (RBM)								& 1.02	& 0.821	& 1.179	& 0.333	\\
\ref{sN_t_vs_mod},\ref{sN_t_vs_nobe},\ref{sN_t_hbb}						& 0.24	& 0.826	& 1.138	& 0.100	\\
\ref{plane_eta},\ref{plane_hbb}									& 1.02	& 0.900	& 1.100	& 0.333	\\
\hline
\end{tabular}
\caption{The dimensions of the annular region divided into bins for $\eta^2$ calculations, in units of $k_0$.
}\label{TableDim}
\end{table}

\begin{figure}[hbt]
  \includegraphics[clip,scale=0.33]{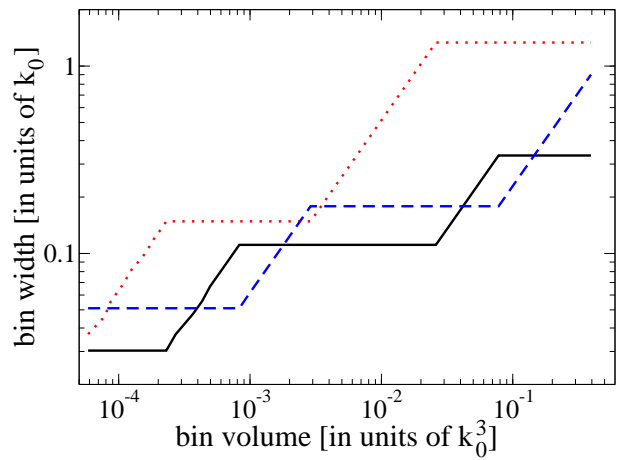}
  \caption{
    (Color online) The bin widths (black solid line -- axial, dotted red line -- tangential, dashed blue line -- radial) as a function of the total bin volume (dense case $\gamma=1.02$) used in the full model
 (\ref{STABeq},\ref{GPeq}) simulations.
  }\label{f7}
\end{figure}

This annular region  is then divided into equi-sized bins by making a series of equally-spaced cuts in the axial (z), radial, and tangential directions.
Then number of cuts is varied to achieve a gradation from large bin sizes covering an angle of $\pi/2$ in the $k_x-k_y$ plane (8 bins in total, 4 pairs),
to bin sizes comparable with the dimensions of a single correlation volume, then on to small bins the size of a single computational lattice point.
The progression of bin dimensions is that we
first reduce the radial bin size to approximately the correlation length $\sigma_r$, then the tangential and axial (z) sizes to their correlation lengths in that order.
Then we repeat reductions in the same order down to single-computational-lattice-point volumes.
One example is shown in Fig.~\ref{f7} for the first $\gamma=1.02$ dense case in the table.

Number squeezing between opposite bins is calculated, then averaged over all pairs to obtain the displayed values of $\eta^2$.
One subtlety should be mentioned: at small bin sizes, mean occupation can vary appreciably as a result of mismatches between bin shapes and the positions of discrete
lattice points on the computational lattice; some small bins miss all lattice points altogether.
Such an effect does not appear experimentally, so to exclude potentially disruptive contributions from atypically discretized bins we exclude some bin pairs from consideration.
The excluded bins are those for which either bin has an ensemble average occupation less than half or more than twice the mean bin occupation (when averaged over all bins).

\section{Calcluation of halo correlations}
\label{APP-G2}

The correlation properties are averaged in a similar way to that performed in experiments \cite{Jaskula10,Kheruntsyan12}.
That is, we calculate the mean correlation function in a region $\mathcal{R}_j$ as a function of inter-particle distance $\Delta k$ in various directions $j$ in the following way:
\begin{equation}
  \bar{g}^{(2)}_{\rm cc/bb}(\Delta k) =
  \frac{\int_{\mathcal{R}_j} d^3\mathbf{k} \langle\dagop{\delta}(\mathbf{k}) \dagop{\delta}(\mathbf{k}+\Delta k)\op{\delta}(\mathbf{k})\op{\delta}(\mathbf{k}+\Delta k)\rangle}
       {\int_{\mathcal{R}_j} d^3\mathbf{k}\ n(\mathbf{k})n(\mathbf{k}+\Delta k)}.
\end{equation}
This region is similar to the annular region used to calculate number-squeezing (as shown in Table~\ref{TableDim} and explained above in Appendix~\ref{APP-BIN}) for the $j=z$ direction,
but a reduced volume for the other directions $j=t,r$, where the reduction consists of additional restriction to within $k_0/3$ of the x and y axes. However, the washer shape is much wider radially, i.e. it has the same average radius as in the table, but a radial width of $0.9k_0$.

The averaging method used here is as employed in experiments because it is convenient for analysis of data consisting of detected particle positions.
It effectively weights the contribution to the correlations proportionally to the product of the densities at the two points $\mathbf{k}$ and $\mathbf{k}+\Delta\mathbf{k}$,
which is approximately the local halo density squared. Thus, it takes into account primarily the most relevant, dense, part of the halo.
Then, Gaussian fits are made to obtain the peak values $h$'s and the widths $\sigma$'s.
During the fitting, points with excessive statistical uncertainty, or for distances $\Delta k_j$ so large that the correlation function $g^{(2)}$ begins to rise, are excluded.

\section{Gaussian model for halo correlations}
\label{APP-eta}
We use Gaussian $g^{(2)}$ functions as in Equations (\ref{g2clgaus}) and (\ref{g2bbgaus}) and approximate the halo by a step function in the radial direction, $k_r = \sqrt{k_x^2+k_y^2}$.
The density is $\rho_0$ when  $|k_r-k_h|<w_h$ and zero otherwise. Here, $k_h$ is the halo mean radius, and $w_h$ the radial halo half-width taken to be $\sqrt{\pi/2}$
times the standard deviation of a Gaussian fit to the true radial profile of the halo density.
We also take the bins to be centered radially at $k_h$ so the effective integration range in the radial direction extends from $-q_r$ to $+q_r$, where $q_r = {\rm min}(L_r/2,w_h)$.
Therefore, the integrals of the correlation functions read
\begin{equation}
  G_{\rm aa/ab} =\rho_0^2\int\limits_{-L_z/2}^{L_z/2}\!\!\!\! dk_z dk'_z\!\!\!  \int\limits_{-L_t/2}^{L_t/2}\!\!\!\! dk_t  dk'_t  \int\limits_{-q_r}^{q_r}\!\! dk_r  dk'_r\,
  g^{(2)}_{\rm cl/bb}(k,k'),\nonumber
\end{equation}
which -- using Equations (\ref{cl}) and (\ref{bb}) -- gives
\begin{equation*}
  G_{\rm aa/ab}= {\bar n}^2 \left[ 1 + h_{\rm cl/bb} f(u_z^{\rm cl/bb}) f(u_t^{\rm cl/bb}) f(u_r^{\rm cl/bb}) \right].
\end{equation*}
Here
\begin{equation*}
  f(u) = \frac{1}{u^2}\left[\sqrt{\frac{\pi}{2}} u\ {\rm erf}(u\sqrt{2}) - \frac{1}{2}\left(1-e^{-2u^2}\right) \right]
\end{equation*}
is a function of the normalized bin widths
\begin{equation*}
  u_{z,t}^{\rm cl/bb} = \frac{L_{z,t}}{2\sigma^{\rm cl/bb}_{z,t}}
  \qquad
  u_r^{\rm cl/bb} = \frac{q_r}{\sigma^{\rm cl/bb}_{z,t}}.
\end{equation*}
This expression is inserted into Eq.~(\ref{eta2avg}) to obtain the estimates of $\eta^2$ on the basis of correlation and density measurements.

\newcommand{\PRL}[1]{Phys. Rev. Lett.~\textbf{#1}}
\newcommand{\PRA}[1]{Phys. Rev.~A~\textbf{#1}}


\begin{thebibliography}{99}

\bibitem{Jaskula10} J.-C.~Jaskula, M.~Bonneau, G.~B.~Partridge, V.~Krachmalnicoff, P.~Deuar, K.~V.~Kheruntsyan, A.~Aspect, D.~Boiron, C.~I.~Westbrook, \PRL{105}, 190402 (2010)
\bibitem{Kheruntsyan12} K.V.~Kheruntsyan, J.-C.~Jaskula, P.~Deuar, M.~Bonneau, G.B.~Partridge, J.~Ruaudel, R.~Lopes, D.~Boiron, C.I.~Westbrook, \PRL{108}, 260401 (2012).
\bibitem{Krachmalnicoff10} V.~Krachmalnicoff, J.-C.~Jaskula, M.~Bonneau, V.~Leung, G.~B.~Partridge, D.~Boiron, C.~I.~Westbrook,
  P.~Deuar, P.~Zi\'n, M.~Trippenbach, K.~V.~Kheruntsyan, \PRL{104}, 150402 (2010).
\bibitem{Vogels02} J.M.~Vogels, K.~Xu, W.~Ketterle, \PRL{89}, 020401 (2002).
\bibitem{Perrin07} A.~Perrin, H.~Chang, V.~Krachmalnicoff, M.~Schellekens, D.~Boiron, A.~Aspect, C.I.~Westbrook, \PRL{99}, 150405 (2007).
\bibitem{Deng99} L.~Deng, E.W.~Hagley, J.~Wen, M.~Trippenbach, Y.~Band, P.S.~Julienne, J.E.~Simsarian, K.~Helmerson, S.L.~Rolston, W.D.~Phillips, Nature \textbf{398}, 218 (1999).
\bibitem{Truscott} Wu RuGway, S. S. Hodgman, R. G. Dall, M. T. Johnsson, and A. G. Truscott, \PRL{107}, 075301 (2011)
\bibitem{MetaStableReview} W. Vassen, C. Cohen-Tannoudji, M. Leduc, ; D. Boiron, C. I. Westbrook, A. Truscott, K. Baldwin, G. Birkl, P. Cancio, M. Trippenbach,
Rev. Mod. Phys. {\bf 84}, 175 (2012)
\bibitem{horo} {R. Horodecki, P. Horodecki, M. Horodecki, and K. Horodecki, Rev. Mod. Phys. {\bf 81}, 865 (2009)}
\bibitem{Reid09} M.D.~Reid, P.D.~Drummond, W.P.~Bowen, E.G.~Cavalcanti, P.H.~Lam, H.A.~Bachor, U.~L.~Andersen, G.~Leuchs, Rev.~Mod.~Phys.~\textbf{81}, 1727 (2009).
\bibitem{kwiat}{P. G. Kwiat, K. Mattle, H. Weinfurter, A. Zeilinger, A. V. Sergienko, and Y. Shih, \PRL{75}, 4337 (1995)}

\bibitem{Kofler12} J. Kofler, M. Singh, M. Ebner, M. Keller, M. Ktyrba, A. Zeilinger, \PRA{86}, 032115 (2012).

\bibitem{giovanetti} {V. Giovanetti, S. Lloyd and L. Maccone, Science {\bf 306}, 1330 (2004)}
\bibitem{pezze}{L. Pezz\'e and A. Smerzi, \PRL{102}, 100401 (2009)}

\bibitem{Bucker11} R. B\"ucker, J. Grond, S. Manz, T. Berrada, T. Betz, C. Koller, U. Hohenester, T. Schumm, A. Perrin and J. Schmiedmayer, Nat. Phys. {\bf 7}, 608 (2011)
\bibitem{Bonneau12} M. Bonneau, J. Ruaudel, R. Lopes, J.-C. Jaskula, A. Aspect, D. Boiron, and C. I. Westbrook, Phys. Rev. A {\bf 87}, 061603(R) (2013).

\bibitem{wine} {D. J. Wineland, J. J. Bollinger, W. M. Itano and D. J. Heinzen, \PRA{50}, 67 (1994)}
\bibitem{kita} {M. Kitagawa and M. Ueda, \PRA{47}, 5138 (1993)}
\bibitem{ita1}{D. J. Wineland, J. J. Bollinger, W. M. Itano, F. L. Moore and D. J. Heinzen, \PRA{46}, 6797 (1992)}

\bibitem{obert}{J. Est\'eve, C. Gross, A. Weller, S. Giovanazzi and M. K. Oberthaler, Nature {\bf 455}, 1216 (2008)}
\bibitem{app}{J. Appel, P. J. Windpassinger, D. Oblak, U. B. Hoff, N. Kj\ae rgaard, and E. S. Polzik, PNAS {\bf 106}, 10960 (2009)}
\bibitem{gross}{C. Gross, T. Zibold, E. Nicklas, J. Esteve and M. K. Oberthaler, Nature {\bf 464}, 1165 (2010)}
\bibitem{riedel}{Max F. Riedel, Pascal Bohi, Yun Li, Theodor W. Hansch, Alice Sinatra and Philipp Treutlein, Nature {\bf 464}, 1170 (2010)}
\bibitem{vule}{I. D. Leroux, M. H. Schleier-Smith, and V. Vuletic, \PRL{104}, 250801 (2010)}
\bibitem{chen}{Zilong Chen, Justin G. Bohnet, Shannon R. Sankar, Jiayan Dai, and James K. Thompson, \PRL{106}, 133601 (2011)}
\bibitem{smerzi} B. L\"ucke,  M. Scherer,  J. Kruse,  L. Pezz\,e, F. Deuretzbacher, P. Hyllus, O. Topic, J. Peise, W. Ertmer, J. Arlt, L. Santos, A. Smerzi and C. Klempt,
  Science {\bf 11}, 773 (2011)

\bibitem{dall} R. G. Dall, L. J. Byron, A. G. Truscott, G. R. Dennis, M. T. Johnsson, and J. J. Hope, \PRA{79}, 011601 (2009)
\bibitem{pertot} D. Pertot, B. Gadway, and D. Schneble, \PRL{104}, 200402 (2010)

\bibitem{Ogren08} M. Ogren, K. V. Kheruntsyan, \PRA{78}, 011602(R) (2008).
\bibitem{Ogren10} M. Ogren, K. V. Kheruntsyan, \PRA{82}, 013641 (2010).

\bibitem{stab} P. ~Deuar, J. ~Chwede\'nczuk, M. Trippenbach, and P. Zin, \PRA{83}, 063625 (2011).
\bibitem{Molmer08} K. Molmer, A. Perrin, V. Krachmalnicoff, V. Leung, D. Boiron, A. Aspect, C. I. Westbrook, \PRA{77}, 033601 (2008).
\bibitem{Bragg} M.~Kozuma, L.~Deng, E.~W.~Hagley, J.~Wen, R.~Lutwak, K.~Helmerson, S.~L.~Rolston, W.~D.~Phillips, \PRL{82}, 871 (1999)


\bibitem{Deuar11epjd} P.~Deuar, P. Zin, J. ~Chwede\'nczuk, and M. Trippenbach, Eur.~Phys.~J.~D {\bf 65}, 19 (2011).

\bibitem{Zin05} P.~Zi\'n, J.~Chwede\'nczuk, A.~Veitia, K. Rz\c{a}\.zewski, M.~Trippenbach, \PRL{94}, 200401 (2005).
\bibitem{Zin06} P.~Zi\'n, J.~Chwede\'nczuk, M.~Trippenbach, \PRA{73}, 033602 (2006).
\bibitem{Chwedenczuk06} J.~Chwede\'nczuk, P.~Zi\'n, K.~Rz\c{a}\.zewski, M.~Trippenbach, \PRL{97}, 170404 (2006).
\bibitem{Chwedenczuk08} J.~Chwede\'nczuk, P.~Zi\'n, M.~Trippenbach, A.~Perrin, V.~Leung, D.~Boiron, C.I.~Westbrook, \PRA{78}, 053605 (2008).

\bibitem{Deuar02} P.~Deuar and P.D.~Drummond, \PRA{66}, 033812 (2002).

\bibitem{pert_Wasak} T. Wasak, J. Chwede\'nczuk, P. Zi\'n and M. Trippenbach, in preparation

\bibitem{Bach02} R.~Bach, M.~Trippenbach, K. Rzazewski, \PRA{65}, 063605 (2002).
\bibitem{Yurovsky02} V.A.~Yurovsky, \PRA{65}, 033605 (2002).
\bibitem{Trippenbach00} M.~Trippenbach, Y.B.~Band, P.S.~Julienne, \PRA{62}, 023608 (2000).
\bibitem{Band01} Y.B.~Band, J.P.~Burke,~Jr., A.~Simoni, P.S.~Julienne, \PRA{64}, 023607 (2001).
\bibitem{Band00} Y.B.~Band, M.~Trippenbach, J.P.~Burke, P.S.~Julienne, \PRL{84}, 5462 (2000).
\bibitem{Ogren09} M.~Ogren, K.V.~Kheruntsyan, \PRA{79}, 021606(R) (2009).
\bibitem{Chwedenczuk04} J.~Chwede\'nczuk, M.~Trippenbach, K.~Rz\c{a}\.zewski, J. Phys. B \textbf{37}, L391 (2004).


\bibitem{Duan00} L.-M. Duan, A. S{\o}rensen, J. I. Cirac, P. Zoller, \PRL{85}, 3991 (2000).
\bibitem{Dennis10} G. R. Dennis, M. T. Johnsson \PRA{82}, 033615 (2010).
\bibitem{Haine11} S. A. Haine, A. J. Ferris \PRA{84}, 043624 (2011).
\bibitem{Williams12} R. A. Williams, L. J. LeBlanc, K. Jimenez-Garcia, M. C. Beeler, A. R. Perry, W. D. Phillips, I. B. Spielman, Science {\bf 335}, 314 (2012).


\end{thebibliography}
\end{document}